\documentstyle[epsfig]{aipproc}
 
\def\Journal#1#2#3#4{{#1} {\bf #2}, #3 (#4)}

\def\ANP{{\em Adv. Nucl. Phys.}}
\def\EPJ{{\em Eur. Phys. J.} A}
\def\APP{{\em Acta Phys. Pol.} B}

\def\NPA{{\em Nucl. Phys.} A}
\def\NPB{{\em Nucl. Phys.} B}
\def\PLB{{\em Phys. Lett.}  B}
\def\PRL{\em Phys. Rev. Lett.}

\def\PRC{{\em Phys. Rev.} C}

\begin{document}
\phantom{.} \hfill NT@UW--99--60 \\
\phantom{..} \hfill DOE/ER/40561--78--INT99

\title{Meson exchange models for Meson production}

\author{C. Hanhart$^*$\protect{\thanks{Supported by USDOE and 
the Alexander-von-Humboldt Foundation}}}
\address{$^*$Department of Physics, University of Washington,
Seattle, WA 98195--1560, USA.}

\maketitle

\begin{abstract}
The production of mesons in nucleon--nucleon collisions is reviewed
from the viewpoint of the meson--exchange picture. 
In the first part
various possible 
production mechanisms and their relative importance are discussed. 
In addition, general features of meson production are described.

In the second part
special emphasis is put on pion production. 
Implications of chiral perturbation theory are discussed.
Results based on a specific
 meson-exchange model for pion production in nucleon--nucleon
collisions are presented.  The model is utilized to calculate several 
spin-dependent observables of the reactions
$pp \to pp\pi^0$ and $pp \to pn\pi^+$
such as spin-correlation parameters.
This allows us to study the role of the Delta resonance at close
to threshold energies.

The talk closes with a brief discussion of the production of 
$\phi$ mesons as an example for the production of heavier mesons.
A strategy is outlined that allows to extract information on the
structure of the nucleon from the reaction $NN \to NN \phi$.
\end{abstract}

\section*{Introduction}

With the advent of new accelerator technology measurements with 
very high accuracy became possible for meson production close to 
 threshold \cite{DATA}. This improvement opens the possibility for the 
investigation of a lot of interesting physics phenomena. It is the 
goal of this talk to highlight a few of those.
This will we done using the meson exchange picture.

The largest part of the presentation will concentrate on pion production. 
As this is the first and most important inelasticity of the 
nucleon--nucleon system it shall deal as a guideline for models 
of the production of heavier mesons. In addition, for this channel
there are not only analyzing powers but also double polarization 
data available. A study of this system will therefore allow us
to get a feeling about what can be learned from meson production.
In particular it will be demonstrated that using polarization observables
will allow to study resonances that couple too weakly to 
significantly influence the unpolarized observables -- in this example
it will be the $\Delta$ (1235) far below the resonance position.
However, the same kind of sensitivity 
should be found for resonances that
have only a minor effect on the
total cross section even at their resonance position.
 In addition: in the unpolarised observables 
almost all the effects of the Delta--isobar are driven by the 
$^1D_2 \to ^5S_2$ $NN \to N\Delta$ transition leading to p--wave
$\pi$--production\footnote{This transition is prohibited for
the $pp \to pp\pi^0$ reaction and this is why the Delta
is less important for the neutral pion production (see below).}. Polarization observables
allow to study less dominant transition amplitudes selectively and 
thus allow direct insight into the NN interaction.
 
There is one special feature to the production of pions: since they are
pseudo Goldstone bosons of the spontaneous broken chiral symmetry,
there is the possibility of studying pion production in 
nucleon--nucleon collisions with chiral perturbation theory.
I will briefly compare the features of the meson exchange 
approach to those of chiral perturbation theory.

As an example for the production of heavier mesons
I will continue with a short discussion of how one can use
data on the production of vector mesons close to threshold to
deduce information on the structure of the nucleon.

The presentation will close with a brief outlook.

\section*{General features of meson production in nucleon--nucleon
collisions}

\begin{figure}[h]
\vspace{4.2cm}
\includegraphics{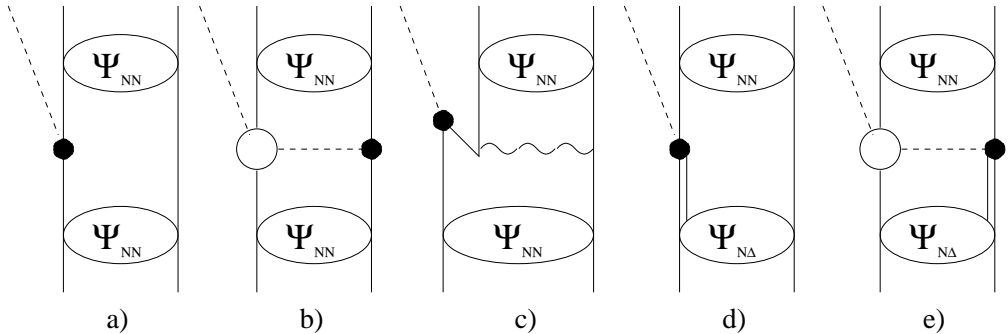}
\caption{Possible pion production mechanisms taken into account in our model:
(a) direct production; (b) pion rescattering; (c) contributions from pair
diagrams; (d) and (e) production involving the excitation of
the $\Delta (1232)$ resonance. Note that diagrams where the
$\Delta$ is excited after pion emission are also included.}
\label{beitraege}
\end{figure}

To produce a meson in nucleon nucleon collisions the initial
kinetic energy needs to be large enough to put the outgoing meson on its mass
shell. The initial center of mass momentum required to produce
a meson of mass $m$ turns out to be at least
$
p = \sqrt{Mm+\frac{m^2}{4}} \ ,
$
where $M$ denotes the nucleon mass.
If the kinematics is chosen close to the threshold the outgoing 
particles are (almost) at rest. Thus, large
characteristic for the meson production close to threshold.
This leads to a big momentum mismatch between the final 
wave-function and the initial one leading to a suppression
of the direct
 production mechanisms (c.f. Fig. \ref{beitraege}a and d for pion production).
Therefore short range processes  dominate the production (c.f. Fig. \ref{beitraege}b
\footnote{For large space like momentum transfers, the $\pi N$ T--matrix enters
far off--shell and thus also the pion exchange contribution should
be considered as a short range process.} and c).
Note, however, once one moves away from threshold
the range of allowed final momenta increases. Thus moderate momentum transfers
become more and more likely and higher partial waves come into play.
Thus, one should expect very
different production mechanisms to be dominant as a function
of the excess energy. 
The range of energies just described can be identified with 
the range of $0 \leq \eta \leq 1$, where $\eta$ denotes the maximum meson
momentum in units of the meson mass. Therefore, in this presentation we
will mainly concentrate on this regime.

The goal of the meson exchange picture is to find a description
of meson production that is both transparent enough to be able
to identify the relevant physics of the process and still contains
all the relevant dynamics.
In this context 
it is important to understand
the special role that the nucleon--nucleon system plays 
in the initial as well as the final state.

As early as 1952 Watson pointed out the significance that the final state
nucleon--nucleon interaction should play in meson production processes
close to threshold \cite{watson}. To see this let us consider  meson 
absorption on a two nucleon system, which is related to  meson production
through detailed balance. The cross section is a measure of the volume
in which all three particles must be simultaneously in order to make
the absorption process possible. However, since the elastic scattering
cross section for the two nucleons is very much larger than the production
cross section and the NN interaction close to threshold is strongly 
attractive, a two step process is favored in which first the two
nucleons approach at close range -- driven by the NN interaction --
before the meson comes into play.
Therefore the energy dependence
of the absorption (production) matrix element should be given by the energy
dependence of the initial (final) nucleon--nucleon interaction.
If we neglect the effect of the meson--nucleon interaction
and use the effective range expansion for the nucleon--nucleon interaction
the energy dependence of the total production cross--section close to threshold
should be given by
\begin{equation}
\sigma_{NN \to NNx}(\eta ) \propto \int_0^{m_x \eta} dq' q'{}^2 
\frac{p'}{1+a(a+r_o)p'{}^2} \ ,
\label{endep}
\end{equation}
where $a$($r_o$) denotes the nucleon--nucleon scattering length
(effective range) in the relevant channel and $p'$ and $q'$
denote the relative momenta of the two nucleons and of the meson
with respect to the two nucleons respectively.
It was shown in ref. \cite{MuS} that the inclusion of the
Coulomb effect is important, if 
for a two proton final state.
The inclusion of the 
Coulomb effect in eq. \ref{endep} 
is straightforward \cite{meyer2}. It turns
out that the energy dependence of the total cross section
that follows from eq. \ref{endep}
is in nice agreement with almost all experimental evidence on 
meson production near
threshold \cite{DATA}. The only exception is the production of $\eta$ mesons, since
the $\eta N$ interaction is too strong to be neglected (see
also ref. \cite{ulfneu}). 

The above argument shows that it is important to include
the final state interactions of the produced particles
 in order to describe the
energy dependence of the total cross sections correctly. Reversing this
statement indicates, that it is possible to relate the energy dependence
of the near threshold production cross sections to low energy parameters
of the dominant interactions. Examples for this are given in ref. 
\cite{janpkl} for the $\Lambda N$ interaction and in refs.
 \cite{Green,Grishina} for the $\eta N$ interaction.

Care should be taken, however, when trying to deduce
a general formula like eq. \ref{endep} not only to
 describe the energy dependence
but also to set the normalization of the effect of the FSI
independent of the production operator.
This statement is illustrated by different groups using 
different prescriptions for the normalization 
of the FSI factor that differ by more than an order of
magnitude, although both use the meson exchange picture to calculate
the production operator (c.f. e.g. refs. \cite{ulfneu} and \cite{sibir2}).
 Based on quite
general arguments one can show, that there is no model independent way
to fix the normalization introduced by the FSI factor \cite{fsicrit}
\footnote{The method applied in refs. \cite{ulfneu,sibir2} was also critizised
in ref. \cite{jounifsi} from a different point of view, namely, that the
inclusion of the FSI through using a plane wave that is modified in strength
according to the on--shell scattering data
misses the short range correlations introduced by the FSI and thus 
leads to wrong relative weights of the individual contributions.}.

A similar reasoning that lead to the successful description of the
energy dependence of the total cross section leads to an expression
that describes the effect of the initial state interaction \cite{fsicrit}:
In order to allow the production to proceed, the two incoming nucleons have to come very close
to each other. There are two mechanisms that potentially prevent this,
namely elastic and inelastic scattering. 
If we focus on the meson exchange picture, it can be argued that the 
principal value integral that introduced the scheme dependence for the FSI
is actually small for high energies\footnote{Remember: there is no such thing as a 
model independent splitting of the NN interaction and the production operator.}. 
Based on this observation one can derive a rough estimate for 
a suppression factor induced by the ISI
(for a discussion of the range of applicability see \cite{fsicrit})
\begin{equation}
\lambda = \frac{1}{4}(1+\eta_L)^2-\eta_L \sin ^2(\delta_L) \ ,
\end{equation}
where $\eta_L$ and $\delta_L$ denote the inelasticity and the phaseshifts
of the partial wave in the initial state relevant for the meson production
under consideration. Using the numbers given by the SAID database \cite{said}
we find for the $\eta$ and $\eta '$ production values of $\lambda = \frac{1}{5}$
and $\lambda = \frac{1}{3}$ respectively. Note, that the former number covers
the bulk effect of the initial state interaction calculated within a meson
exchange model \cite{lee}. Thus, the ISI leads to a non negligible suppression
of the total cross section and should be taken into account.

\section*{Pion production and the role of the $\Delta$ close to threshold}

As this is meant to be an overview presentation, this section starts
with a brief history of the field.

Essentially all recent theoretical investigations on pion production
near threshold are based on the model proposed by
Koltun and Reitan in 1966 \cite{KuR}. In this model two production
mechanisms are considered:  direct production
(Fig.~\ref{beitraege}a), and pion rescattering (Fig.~\ref{beitraege}b) --
the latter, however, in on--shell approximation, where the $\pi$N--T--matrix
is replaced by the scattering length. It is important to note that --
as a consequence of chiral symmetry -- the isoscalar scattering
length\footnote{The $\pi$N--T--matrix can be written as a linear combination
of an isoscalar and an isovector part.} almost vanishes.
This model was utilized by Miller and Sauer in 1991 \cite{MuS}
to analyze the first set of high precision data of the
reaction $pp \rightarrow pp\pi^0$ near threshold that became available
from IUCF \cite{IU1}.
It turned out that such a model grossly underestimates
the empirical cross section \cite{MuS}. Note, that isovector rescattering
is prohibited in this reaction channel.

In 1992 Niskanen extended this model by including the $\Delta$ (1235)
isobar (cf. Fig.~\ref{beitraege}d and e) \cite{Nis1}. Furthermore, he allowed for
an energy dependence in the s-wave rescattering term, however,
still keeping the on--shell approximation \cite{Nis2}. These
improvements roughly doubled the predicted $pp\rightarrow pp\pi^0$
cross section, but Niskanen's results still underestimate the IUCF data
by a factor of 3.6.

Another new production mechanism was introduced by Lee and Riska
in 1993 \cite{LuR}. These authors considered effects from meson-exchange
currents due to the exchange of heavy mesons
that excite a nucleon--antinucleon pair, as shown in Fig.~\ref{beitraege}c.
It was found that the resulting contributions (in particular the
one of the $\omega$ meson) enhance the pion production cross section
by a factor of 3-5 \cite{LuR,HMG} and thus eliminate most of the
under prediction found in earlier investigations.

However, in 1995 Hern\'andez and Oset presented an alternative
explanation for the missing strength in the $\pi^0$ production close
to threshold \cite{HO}. These authors took into account the
off-shell properties of the $\pi N$ amplitude in the
evaluation of the rescattering diagram.
It turned out, that this enhances the rescattering contribution sufficiently
to describe the empirical $pp \to pp\pi^0$ cross--section without
the inclusion of additional short ranged diagrams.
In the same year our group performed a calculation using a microscopic
model for the $\pi$N T--matrix \cite{Han1}. Although the effect
of the rescattering was not as large as reported in \cite{HO} its 
enhancement due to the off--shell part of the T--matrix was quite significant.

\begin{figure}[t]
\vspace{9cm}
\includegraphics{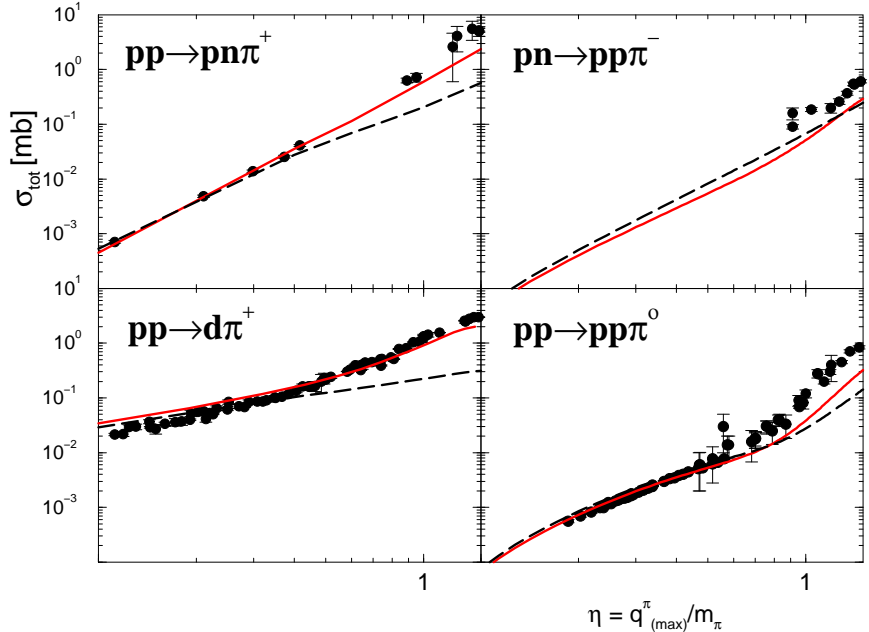}
\caption{\it{Total cross sections for the different
pion production channels.
The solid line represents the result
of the  full model of ref. \protect\cite{Han98a}, whereas for the dashed line the contributions
involving the $\Delta$ where omitted. 
For a compilation of the experimental data see ref. 
\protect\cite{DATA}.
}}
\label{csresults}
\end{figure}

Since the pion is the Goldstone Boson of the chiral symmetry one might
hope that the framework of chiral perturbation theory allows more insight
into this fundamental reaction. In the literature there is a series 
of papers available, that use tree level chiral perturbation theory to calculate
the cross sections close to threshold for both neutral pion production
\cite{CP1,CP2,CP2b,CP3} and charged pion production \cite{HHH,CP4}.
In addition there are two calculations available that calculate the
$\pi^o$ production to one loop order \cite{moalem,fred} -- these, however,
use heavy baryon chiral perturbation theory which is inappropriate in
the present kinematics \cite{ulfneu}. 
The advantage of chiral perturbation theory compared to the meson exchange
picture is, that one has an organizing principle at hand. Although the 
expansion parameter is quite large in case of the $NN \to NN\pi$ reactions,
namely $Q=\sqrt{m/M} \simeq \frac{1}{3}$ \cite{CP2}, 
it is very illuminating to study the 
lowest order contributions and to compare those to the diagrams depicted
in Fig. \ref{beitraege}.

Based on the counting rules employed in ref. \cite{CP2} the lowest order
contributions to pion production close to threshold (O($Q$)) are for the neutral
pion production the direct production from the nucleon as well as
from the delta (Fig. \ref{beitraege}a and d), whereas in case of the
charged pion production at the same order there appears a rescattering
diagram (Fig. \ref{beitraege}b) involving the isovector $\pi N$ interaction
in addition \cite{HHH,CP4}.
As it was argued above, the large momentum transfer characteristic of 
meson production close to threshold leads to a suppression of the
direct production mechanisms in favor of rescattering and short 
range processes. Therefore, one expects the production of charged pions to 
be under better control than the one of neutral pions -- that is 
indeed what we experience: the lowest order calculation for the $\pi^+$ production
deviates from the data by a factor of 2 \cite{HHH,CP4} whereas the one for $\pi^0$ production
deviates by more than an order of magnitude (the latter occurs because the 
individual contributions are small and there is a cancelation between
the $\Delta$ contribution (Fig. \ref{beitraege}d) and the nucleonic one (Fig. \ref{beitraege}a)
 \cite{CP2}).

\begin{figure}[t]
\vspace{10cm}
\includegraphics{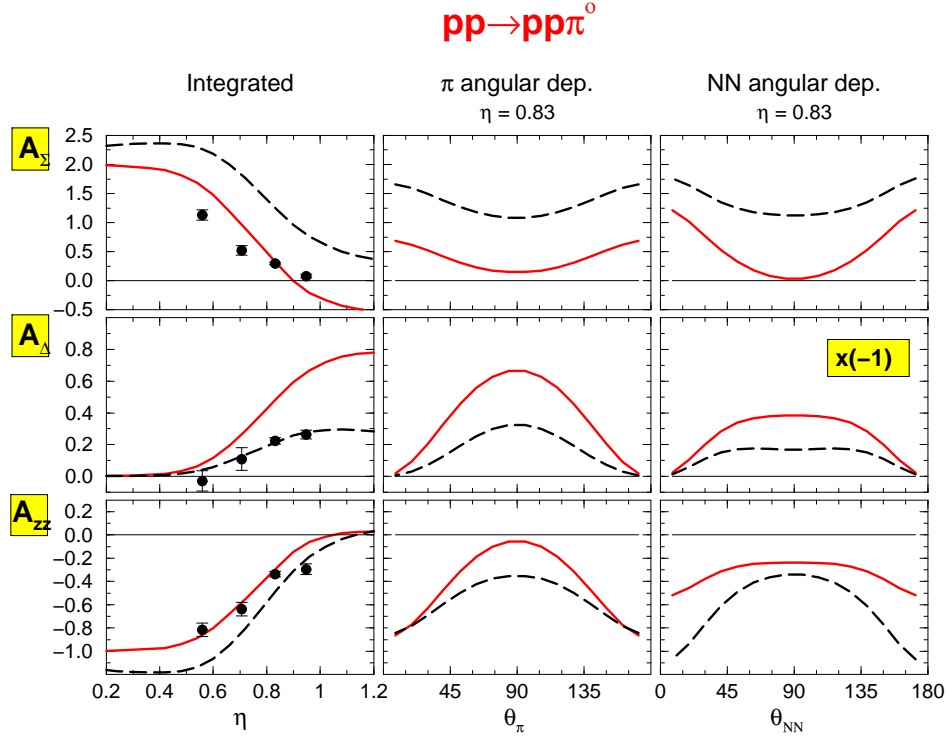}
\caption{\it{Spin correlation parameters for the reaction
$pp \to pp\pi^0$, where $A_\Sigma = A_{xx}+A_{yy}$ and
$A_\Delta = A_{xx}-A_{yy}$. 
Curves are the same as in Fig. \protect\ref{csresults}.
The experimental data are taken from 
Refs.~\protect\cite{Mey98,Mey99}.
}}
\label{Corr1}
\end{figure}

At the next order (O($Q^3$)) in both channels loops enter\footnote{This
is true only when using the counting scheme of ref. \cite{CP2}.} as well
as contact interactions (the effective field theory
analog of the short range contributions of the meson exchange picture) and
 rescattering diagrams. The strength of the contact interactions
is not constrained by chiral symmetry. Therefore the authors of refs. 
\cite{CP2,CP2b,CP4} tried to estimate their strength by means of resonance
saturation: the contact interactions were identified with short ranged
diagrams (Fig. \ref{beitraege}c) that were evaluated using parameters
from realistic $NN$ potentials. However, it is not clear a priori, that
this procedure makes sense for the contact interactions.
The parameters relevant for the rescattering can be related to 
$\pi N$ scattering \cite{bkm}. 

Taking all this together it should not come as a surprise that the s--wave pion
production is not well under control theoretically. As it was argued 
before the typical momentum transfers decrease as we move away from the
production threshold and it is the large momentum transfer that causes all
the trouble. 
A more appropriate approach to pion production close to threshold thus seems
to be to first understand the higher partial waves and then to move down 
in energy. This procedure should allow one to study the onset of new 
physics in a more controlled way what
highlights the importance of looking at higher partial
waves. It should be noted that 
one expects 
 the chiral expansion to work better, when the pions are produced in  higher
partial waves\cite{biraprivate}. Therefore in what follows we will concentrate on polarization
observables for here the higher partial waves show up most clearly.

To be more concrete in what follows we will compare the results of a particular
meson exchange model \cite{Han98a} to the data.
 In the literature there is only one additional
model that considers all the different pion production channels as well as
higher partial waves, namely the one by the Osaka group.
This model was described in a different presentation in some detail  \cite{Tamura},
and thus we will not discuss it here.

In the J\"ulich model \cite{Han98a} all standard pion-production mechanisms
(direct production (Fig.~\ref{beitraege}a),
pion rescattering (Fig.~\ref{beitraege}b),
contributions from pair diagrams (Fig.~\ref{beitraege}c)
are considered -- the latter, however, are treated
as a parametrisation of all the missing short range mechanisms.
In addition, production mechanisms involving the
excitation of the $\Delta (1232)$ resonance 
(cf. Fig.~\ref{beitraege}d,e) are taken into account explicitly. 
All $NN$ partial waves up to orbital angular momenta $L_{NN} = 2$, and 
all states with relative orbital angular momentum $l \leq 2$ between 
the $NN$ system and the pion are considered in the final state. 
Furthermore all $\pi N$ partial waves up to orbital angular momenta
$L_{\pi N} = 1$ are included in calculating the rescattering diagrams
in Fig.~\ref{beitraege}b,e. Thus, our model includes not only
s-wave pion rescattering but also contributions from 
non resonant p-wave rescattering.
For details about the model we refer the reader to ref. \cite{Han98a}.

\begin{figure}[t]
\vspace{10cm}
\includegraphics{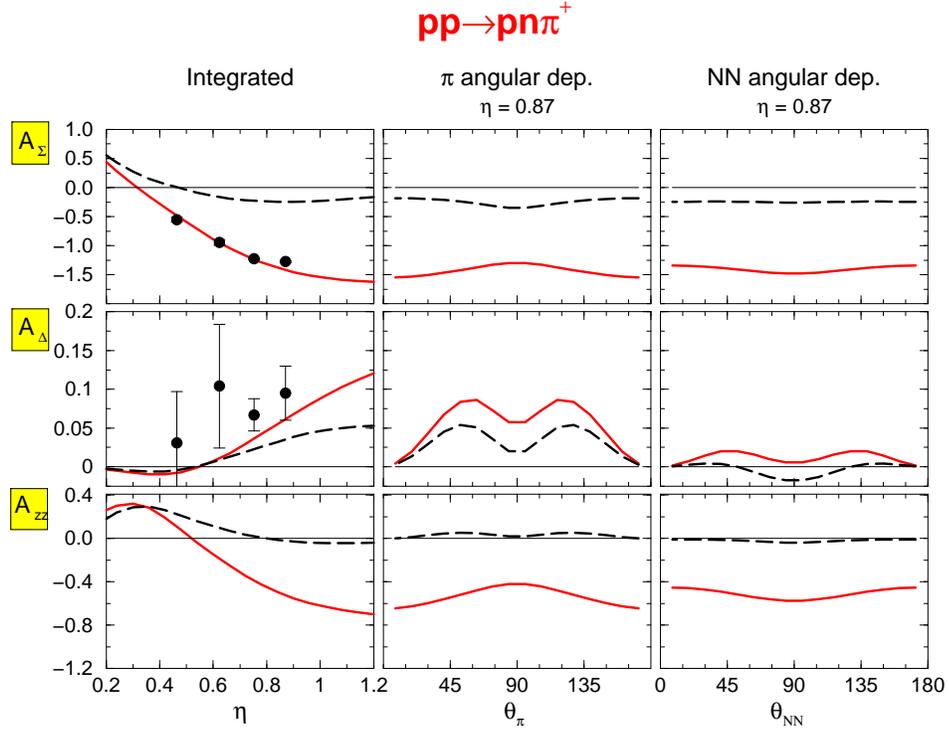}
\caption{\it{Spin correlation parameters for the reaction
$pp \to pn\pi^+$. Same description of the curves as in
Fig.~\ref{csresults}. The experimental data are taken from 
Ref.~\protect\cite{Sah99}.
}}
\label{Corr2}
\end{figure}

Results of this model for total cross sections for the reactions channels
$pp \rightarrow pp\pi^0$, $pp \rightarrow pn\pi^+$,
$pn \rightarrow pp\pi^-$, and $pp \rightarrow d\pi^+$ are displayed in 
Fig. \ref{csresults}.
Results for
 analyzing powers
 were
presented in Refs.~\cite{Han98a,Han98b}. It was found that the model 
yields a very good overall description of the data from the 
threshold up to the $\Delta$ resonance region. 
In fact, a nice quantitative agreement with basically all experimental
information (then available) was observed over a wide energy range.
Thus, this model is very well suited as a starting point for a detailed
analysis of the forthcoming spin-dependent observables of the reaction 
$NN \rightarrow NN\pi$.

Looking at the results for the total cross sections one observes
that the role of the $\Delta$ is very different in the different
production channels: it dominates the $\pi^+$ production at 
values of $\eta > 1$ but it appears much less relevant anywhere else.
As we will see in the following, this is no longer true for
polarization observables. Here the
$\Delta$ turns out to be very important in all 
production channels in the whole energy range
if one wants to describe the spin observables as well.
This clearly demonstrates the power of polarization
observables to investigate resonances even if their effect
is too weak to show up in unpolarized observables!

Predictions for the spin correlation coefficient combinations 
$A_\Sigma = A_{xx}+A_{yy}$, $A_\Delta = A_{xx}-A_{yy}$ and $A_{zz}$ 
are shown in Figs.~\ref{Corr1} (for $pp\rightarrow pp\pi^0$)
and \ref{Corr2} (for $pp\rightarrow pn\pi^+$). The polar integrals of 
these observables are displayed in the left panels as a function of 
$\eta$.
The other two panels contain the results for the different observables 
as a function of the pion angle (middle
panel) and of the angle between the nucleons (right panel). 
Data for these observables are currently being analyzed \cite{jan}.

Again the solid line shows the results for the full model, whereas the dashed line
are the results when all contributions involving the $\Delta$ are omitted.
As can be seen clearly from the figure, the inclusion of the isobar is
essential for all the polarization observables displayed even at energies
as low as $\eta = 0.4$!

Our model is not able to describe $A_{xx}-A_{yy}$ in the reaction $pp \to pp
\pi^0$ (c.f. Fig. \ref{Corr1}). The numerator of this observable is sensitive
to $Pp$ partial waves only \cite{Meyba} (Here capital letters denote the NN relative
 momentum whereas the small letters denote 
the pion angular momentum with respect to the NN system).
Does figure \ref{Corr1} therefore tell
us that all but the $Pp$ piece is correctly reproduced by our model\footnote{This
was given as a possible explanation of this deviation in ref. \cite{Han98a}.}?
The answer is no. There is a possible other explanation, namely that it is not
the numerator of  $A_{xx}-A_{yy}$ but the denominator that causes the deviation.
Note, that our model underestimates the total cross section for $\pi^0$ production
by a factor of 2 at energies of $\eta = 1$ (c.f. Fig. \ref{csresults}).

%

As it was pointed out in ref. \cite{Mey99}, it is possible
to relate the integrated double polarization observables
to the spin cross sections (the cross sections one gets
for a given initial spin state). At moderate energies
these should be dominated by the lowest partial waves.
 A possible way to get better 
insights could therefore be to look at the spin cross sections directly. However,
in their determination the total cross section enters besides the 
double polarization observables.
The former, however, is badly known at energies around $\eta = 1$.
Better data is needed to 
accurately constrain the spin cross sections in the energy region of interest.
Fortunately those data will soon be available \cite{Jozef}.

\section*{Vector meson production}

In this section we will briefly discuss, how the investigation of vector meson
production in nucleon--nucleon collisions can improve our knowledge on
the structure of the nucleon. In particular we will outline a method
on how to deduce information on the $NN\phi$ coupling constant from
data on differential cross sections recently measured by the DISTO collaboration \cite{Disto}. 

This information can be regarded as complementary to 
recent experiments on $\bar pp$ annihilation at rest, where 
the unexpectedly large cross section 
ratios $\sigma_{\bar pp \rightarrow \phi X} / 
\sigma_{\bar pp \rightarrow \omega X}$
(cf. Ref.\cite{Ell95} for a compilation of data)
were interpreted by some 
authors as a clear signal for an intrinsic 
$\bar ss$ component in the nucleon\cite{Ell95}. However, in 
an alternative approach based on two-step processes these data
were explained without introducing any strangeness 
in the nucleon and any explicit violation of the OZI rule\cite{Loche}.

In this context $\phi$ production in nucleon-nucleon collisions is
of specific interest. Here one does not expect any significant
contributions from competing OZI-allowed two-step mechanisms.
Therefore cross section ratios 
$\sigma_{pp\rightarrow pp \phi} / \sigma_{pp\rightarrow pp \omega}$
should provide a clear indication for a possible OZI violation and 
the amount of hidden strangeness in the nucleon -- note, that
data recently presented by the DISTO collaboration indicate that this
ratio is about 8 times larger than the OZI estimate\cite{Disto}. 
Based on a model calculation in this section it will be investigated, if the 
observed enhancement over the OZI estimate in the cross section implies
a $g_{NN\phi}$ that is likewise enhanced and therefore 
at variance with the OZI rule. It should be emphasized that the investigation
is by far not as ambitious as the study of pion production, for the
individual ingredients are much less known. The major goal of
this study is to extract constraints on $g_{NN\phi}$ from the $\omega$ to $\phi$ ratio.

As before 
we also  describe the $pp \rightarrow pp\phi$ reaction within a 
relativistic meson-exchange model, where 
the transition amplitude is calculated in DWBA
in order to take the $NN$ final state interaction into account. (See 
Ref.\cite{Nak1} for the details of the formalism.) 
For the $NN$ interaction we employ the model Bonn B\cite{Mach89}. 
The  effect of the ISI is accounted for via an appropriate adjustment of the 
(phenomenological) form factors at the hadronic vertices.

\begin{figure}[t]
\vspace{4.0cm}
\includegraphics{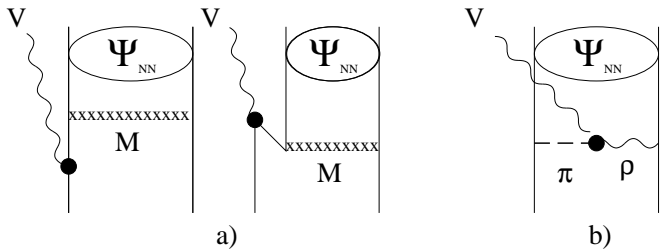}
\caption{$\phi$ and $\omega$-meson production currents included in 
the present study: (a) nucleonic current (the diagrams with
the production after the meson exchange are also included),
 (b) meson exchange current. 
$v = \omega, \phi$ and $M = \pi, \eta, \rho, \omega, \sigma, a_o$.}
\label{Fig1}
\end{figure}

In a previous study of the reaction $pp\rightarrow pp\omega$\cite{Nak1}
it was found that the dominant production mechanisms are the nucleonic and 
$\omega\rho\pi$ mesonic currents, as depicted in Fig.~\ref{Fig1}. 
Note, that in contrast to the pion production here we take the
Born diagram for the $\pi N \to \omega N$ T--matrix multiplied with a 
phenomenological form--factor and not a microscopic T--matrix.
However, this is not a serious draw back for the current investigation,
since this formfactor is anyhow adjusted to the data.
The reason why it is still possible to extract important information
from the vector meson production is,
that the angular distribution of the produced vector
meson provides a unique and clear signature of the magnitude of these
currents, thus allowing one to disentangle these two reaction mechanisms.
Therefore, it is possible
to fix uniquely the magnitudes of the nucleonic and the meson-exchange 
current by analyzing the angular
distribution of the $\phi$ meson measured by the DISTO 
collaboration\cite{Disto}.
Furthermore, since the $NN\phi$ coupling constant enters only in the 
nucleonic current it is possible to extract its value from such an
analysis. It is determined by the requirement of getting the 
proper contribution of the nucleonic current needed to reproduce the 
angular distributions.  

The parameters of the model (coupling constants, cutoff masses of the
vertex form factors)
are mostly taken over from the employed $NN$ model. The $\phi\rho\pi$ 
coupling constant is obtained from the measured decay width of 
$\phi \rightarrow \rho + \pi$. However, besides the $g_{NN\phi}$ that we 
want to 
extract from the analysis, there are still some more free parameters:
The cutoff mass of the $\phi\rho\pi$ vertex form factor
of the meson-exchange current, and the form factor and tensor- to 
vector coupling constant ratio $\kappa_\phi \equiv f_{NN\phi}/g_{NN\phi}$ 
of the nucleonic current. It is possible to fix most of them by 
performing a combined analysis of the $\omega$ and $\phi$ data
as well as other sources
 (cf. Ref.\cite{Nak2} for details). 
The remaining unknown is the tensor coupling. Here
we assume that $\kappa_\phi = 
\kappa_\omega$, as also suggested by SU(3) symmetry,
and $-0.5 \leq \kappa_\omega \leq 0.5$.

\begin{figure}[t]
\vspace{7.3cm}
\includegraphics{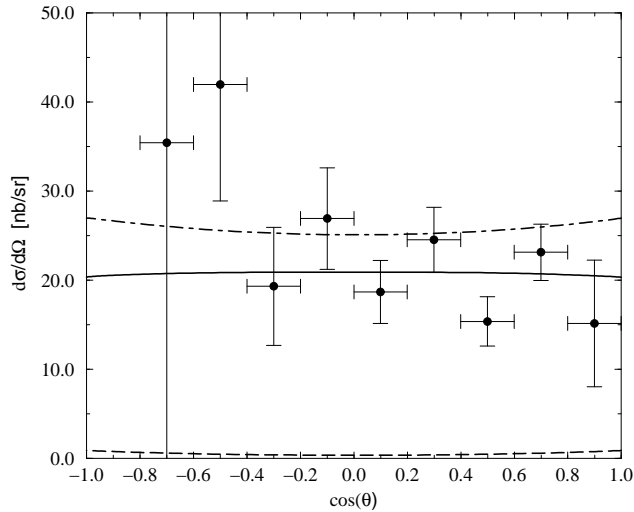}
\caption{Angular distribution for the reaction $pp\rightarrow pp\phi$ 
at an incident energy of $T_{lab}=2.85$ GeV. The dashed-dotted curve corresponds 
to the mesonic current contribution, the dashed curve to the nucleonic 
current contribution. The solid curve is the total contribution. The experimental 
data are from Ref.~\protect\cite{Disto}.}
\label{Fig2}
\end{figure}

After the above considerations, we are now prepared to apply the model 
to the reactions $pp\rightarrow pp\omega$ and $pp\rightarrow pp\phi$.
The angular distribution for $\phi$-meson production measured at 
$T_{lab} = 2.85$ GeV is shown in Fig.~\ref{Fig2}. 
We observe that the angular distribution is fairly flat. Recalling the 
results we obtained for $\omega$ production\cite{Nak1} this 
tells us that $\phi$-meson production should be almost entirely due to the 
$\phi\rho\pi$ meson-exchange current. Only a very small contribution 
of the nucleonic current is required if the angular 
distribution drops at forward and backward angles, as indicated by
the data\footnote{Note, that since there are identical particles in
the initial state the angular distribution has to be symmetric around
90 degrees.}. 

Details about our strategy 
for fixing the various parameters and information about the
data used in the analysis can be found 
in Ref.\cite{Nak2}. 
We get  a set of values 
which range from 
\begin{equation}
-1.4 \ \leq  \ g_{NN\phi}^{model} \ \leq \ -0.163 \ .
\end{equation}
Note, 
that the spread is not an estimate of the theoretical uncertainty of the model
but reflects the remaining uncertainty within the model -- this will be
reduced sufficiently as soon as more experimental information is available.
Nevertheless, it is encouraging to see that the extracted values all lie within
fairly narrow bounds. This clearly indicates to us that the dependence on
the model parameters is not very strong, and that the magnitude of 
$g_{NN\phi}$ is primarily determined by the experimental information used. 

The values of $g_{NN\phi}$ obtained may be compared with those
resulting from SU(3) flavor symmetry considerations and imposition of the 
OZI rule,  
$$
g_{NN\phi}^{OZI} = - 3 g_{NN\rho} \sin(\alpha_v) \cong  -(0.60 \pm 0.15) \ ,
$$
where the factor $\sin(\alpha_v)$ is due to the deviation from the ideal
$\omega - \phi$ mixing. The numerical value is obtained using the values 
of $g_{NN\rho} = 2.63 - 3.36$\cite{rhocoup} and $\alpha_v \cong 3.8^o$. 
Comparing this value with the ones extracted from our model analysis, we 
conclude that the preliminary data presently available can be described with 
using a $NN\phi$ coupling constant that is compatible with the OZI rule.
This clearly indicates that a dynamical model is needed for drawing
any conclusion about the validity of the OZI rule.

\section*{Summary and Outlook}
The main points of the presentation can be summarized as
\begin{itemize}
\item it is important to take into account the nucleon--nucleon interaction
in the final as well as in the initial state
to get a quantitative understanding of meson production close to threshold,
\item the interesting energy regime for meson production is $0 \leq \eta \leq 1$, for 
this regime includes both large and moderate momentum transfers,
\item polarization observables allow to detect resonances, even if their effect
is too small to show up in the total cross section. Note, that this statement
is true only if there is a spin dependent coupling of the resonance to the produced
meson.
\end{itemize}
It should be made clear that theory lacks far behind experiment. However, 
the large amount of experimental information that is going to be available within
the next years will help us to get a better understanding of meson production
reactions and will therefore guide us to better models and insights into
the phenomenology of the nucleon--nucleon interaction.

\section*{Acknowledgments}

The author is very grateful to J. Haidenbauer for comments on the manuscript 
and to G. A. Miller and U. van Kolck for useful discussions.
He also thanks K. Nakayama, J. W. Durso, J. Haidenbauer,
 J. Speth and O. Krehl for the collaboration that lead to the results presented here.

\end{document}